\documentclass[11pt]{amsart}
\usepackage{geometry}                
\usepackage{graphicx}
\usepackage{amssymb}
\usepackage{epstopdf}
\title{Unitary Interactions Do Not Yield Outcomes:
Attempting to Model ``Wigner's Friend''}

\begin{document}
\maketitle

\centerline{\small R. E. Kastner, University of Maryland, College Park. rkastner@umd.edu}
\centerline{\small Version of March 28, 2023; published in Foundations of Physics (2021)51: 89}   
\bigskip

ABSTRACT. An experiment by Proietti {\it et al} purporting to instantiate the `Wigner's Friend' thought experiment is discussed. It is pointed out that the stated implications of the experiment regarding the alleged irreconcilability of facts attributed to different observers warrant critical review. In particular, violation of a Clauser-Horne-Shimony inequality by the experimental data actually shows that the attribution of measurement outcomes to the ``Friends'' (modeled by internal photons undergoing unitary interactions) is erroneous. An elementary but often overlooked result regarding improper mixtures is adduced in support of this assessment. A counterexample is provided which refutes the popular notion that quantum theory leads to `relative facts' that never manifest as empirical inconsistencies. It is further noted that under an assumption of unbroken unitarity, no measurement correlation can ever yield an outcome, since all systems remain in improper mixtures, and attributing a definite but unknown outcome contradicts their composite pure state. It is pointed out that there already exists a solution to this conundrum in the form of an alternative formulation of quantum theory, which accounts for the data showing that no outcomes occurred at the interior entangled photon level and also predicts that outcomes can and do occur at the exterior ``super-observer'' level in this type of experiment. 
\bigskip

\section{Background.}

This paper presents a critique of a recent attempt by Proietti {\it et al} (2019) to experimentally model the Wigner's Friend experiment (WF), a famous elaboration by Wigner of the Schr\"odinger's Cat experiment (SC). While the present work addresses the specific details of that experiment, the critique applies more broadly to all recent discussions of the WF experiment that assume that outcomes occur for subsystems of an entangled state; i.e., that outcomes can ``coexist'' for the ``friend'' who is modeled as a subsystem of an entangled state, as well as for the external ``Wigner'' level. These two levels have been termed in the literature (such as in the scenario of Frauchiger-Renner,  2018) as the ``observer'' and ``super-observer'' level, respectively.  Such a treatment is also found for example in Bong {\it et al} (2020).

Both SC and WF are thought experiments that amplify the measurement problem (MP) of standard quantum theory by showing that it leads to absurdities when entanglements are presumed to continue to the macroscopic level of everyday experience. This author has argued elsewhere (Kastner, 2020b) that these thought experiments are properly understood not as experiments proposed to be performed in the lab (since what they predict, such as cats or people in superposition, is already known to be falsified by experience), but rather as {\it reductio ad absurdum} arguments making explicit the arbitrariness of the so-called `shifty split' that is inevitable when quantum theory is assumed to have only unitary (linear) dynamics. In what follows, I will refer to this standard approach to quantum theory as `UOQM' for `Unitary-Only Quantum Mechanics.'  UOQM consists of any approach that assumes that quantum theory has only unitary physical processes except perhaps upon `observation,' where that term is ambiguous. In addition, UOQM assumes that any non-unitary `collapse' or `reduction' upon `observation' has no quantitative physical counterpart in the theory, making it an {\it ad hoc} postulate. This is the approach assumed in Poietti {\it et al} and in other works claiming that outcomes occurred at the ``friend'' level, at least implicitly.

Before dealing with the specifics of the experiment (which is quite an impressive technical accomplishment despite its misinterpretation), it is important to review an elementary but often-overlooked fact regarding the fraught notion of ``measurement'' in the context of UOQM. This fact is that, as a purely logical matter, there can be no well-defined measurement outcomes under UOQM. This was originally pointed out by Feyerabend (1962) and instructively reviewed by Hughes (1989),\textsection 5.8. We now review the basic argument.

Under UOQM, a good (sharp) measurement of some observable A on a system S is modeled as an interaction between a measuring apparatus or ``pointer'' P, initially in a ready state $|\phi_0\rangle $, that couples it to S such that states $|\phi_i\rangle$ of P point to the different eigenstates $|a_i\rangle$ of A, as follows:

$$ |\phi_0\rangle \otimes |\psi\rangle \rightarrow  c_1 |\phi_1\rangle |a_1\rangle + c_2 |\phi_2\rangle |a_2\rangle
+ c_3 |\phi_3\rangle |a_3\rangle + \ldots c_N |\phi_N\rangle |a_N\rangle  \eqno(1)$$ \medskip

\noindent where the $c_i$ are the amplitudes associated with this decomposition of the system's prepared state  $|\psi\rangle$. Thus, the pointer P enters into an entanglement with S such that states of P ``copy'' the eigenstates of $|a_i\rangle$ pertaining to the system, with the appropriate amplitudes. However, according to standard UOQM, nothing beyond this unitary process ever physically happens--at least nothing that can be modeled in any quantitative way. Either (i) a ``projection postulate'' (PP) is invoked based on an undefined notion of ``observer,'' or (ii) the entanglement is simply assumed to continue. Under (i), one cannot define the physical circumstances constituting an ``observer'' or ``observation''; this is the ``Heisenberg Cut'' or ``shifty split'' in which certain systems are taken as described by quantum theory while others are not, for no principled reason. Under (ii), one cannot say that any outcome actually occurred, at least not one in which the system can be taken as represented by the corresponding eigenstate, even if we don't know which one it is. 
 
The reason is as follows. For simplicity, let us suppose that the observable A has only two eigenstates (like the observables in the Proietti {\it et al} experiment), i.e., the composite state $|\Psi\rangle$
is:

$$|\Psi\rangle = c_1 |\phi_1 , a_1\rangle + c_2 |\phi_2 , a_2\rangle \eqno(2) $$
\smallskip

\noindent  where we use abbreviated notation for the direct product.

The state of the system S, as a component of the entangled pure state (2), is given by the reduced density matrix $\rho_S$. This is obtained by tracing the composite density matrix $\rho$ over the pointer basis.  
In this case, the composite pure state density matrix $\rho$ is:

$$\rho = |c_1|^2 | \phi_1, a_1\rangle\langle \phi_1, a_1| + c_2^*c_1  | \phi_1, a_1\rangle\langle \phi_2, a_2|
+ c_1^*c_2 | \phi_2, a_2\rangle\langle \phi_1, a_1| + |c_2|^2 |\phi_2,a_2\rangle\langle \phi_2, a_2 |  \eqno(3)   $$
\medskip

\noindent and the reduced density matrix for the system is

$$\rho_S = |c_1|^2 |a_1\rangle\langle a_1| + |c_2|^2 |a_2\rangle\langle a_2| \eqno (4)$$

\medskip
This looks like a mixed state in which outcomes occur with the relevant Born probabilities. However, it is an improper mixed state, meaning that it cannot represent a situation in which P and S really possess an outcome represented by the eigenstates  $|\phi_k\rangle$ and $|a_k\rangle$ while the probabilities just represent our ignorance about which outcome has occurred. For were that the case, P and S would be in the mixed state  

$$\rho = |c_1|^2 | \phi_1, a_1\rangle\langle \phi_1, a_1| +  |c_2|^2 |\phi_2,a_2\rangle\langle \phi_2, a_2 |  \eqno(5)   $$

\bigskip

\noindent That is, the off-diagonal entries, reflecting interference obtaining in the pure state (3), are missing. Since the assumption that the system and pointer are really in well-defined eigenstates corresponding to an outcome contradicts the actual composite state, that assumption is false: a unitary interaction yielding a correlation between degrees of freedom, in which one ``copies'' the other, does not constitute the occurrence of a well-defined outcome. With this point in mind, we now turn to the specifics of the experiment.
\bigskip

\section{The Experiment}

The experiment uses a source $S_0$ of Bell-entangled photons $a$ and $b$ that are prepared in the state

 $$ {| \bar \Psi  \rangle}_{ab} \rangle = 
 \frac{1}{\sqrt2} [cos \frac{\pi}{8} (|h_a,  v_b\rangle + |v_a, h_b\rangle)
 + sin \frac{\pi}{8} (|h_a, h_b\rangle - | v_a, v_b \rangle )]   \eqno(6)  $$

\smallskip
\noindent where $|h_{a/b}\rangle$ stands for ``a/b is horizontally polarized'' and $|v_{a/b}\rangle$ stands for ``a/b is vertically polarized.''
The setup is schematically illustrated in Figure 2 of Proietti {\it et al} (2019). The photons $a$ and $b$ are sent in opposite directions, to be ultimately detected by Alice and Bob respectively. Alice and Bob have a choice of measuring either a Bell-type observable $A_1$ ($B_1$) or a ``which polarization'' observable $A_0$ ($B_0$) by leaving in or removing a final beamsplitter. Between the source and the Alice/Bob detectors on each side are two additional sources of entangled photon pairs, $S_A$ and $S_B$. We now describe the situation on the ``Alice'' side; the same interactions occur on the ``Bob'' side. Source $S_A$ produces an entangled pair $\{\alpha, \alpha^{\prime}\}$ in the state  

$$|\Psi ^- \rangle =  \frac {1}{\sqrt 2} [  |h\rangle _{\alpha^\prime}
 |v\rangle _\alpha -    |v\rangle _{\alpha^\prime} \rangle |h\rangle _{\alpha}]  \eqno(7) $$
 
 \smallskip 
\noindent while $S_B$ produces the same state for $\{\beta, \beta^{\prime}\}$. Here, the photons $\alpha$ and $\beta$ are supposed to play the role of ``Alice's Friend'' and ``Bob's Friend'' respectively while  $\alpha^\prime$ and $\beta^\prime$ are alleged to ``herald'' the measurement purported to take place. In follows, we will refer to $\alpha$ and $\beta$ collectively as ``Photon Friends*,'' or PF*s for short, where the asterisk denotes some question as to whether they really qualify as ``Friends'' in the sense of the original Wigner's Friend experiment.  The putative ``measurement'' by the PF*s consists of the establishment of a correlation between  $\alpha$ and $a$ and $\beta$ and $b$, which will be discussed in detail below. When the auxiliary ``heralding'' photons $\alpha^\prime$ and $\beta^\prime$ are detected by a neutral detector, this signals that the correlation has been established. 

\bigskip
\section{Critique of Interpretation of the Experiment}

The first thing to note about the authors' portrayal of the PF*s'  ``measurement'' interactions with the source photons {\it a} and {\it b} is that they misleadingly suppress the fact that the latter are in an entangled state. Their interaction with the PF*s then just increases the entanglement by correlating  $\alpha$ with $a$ and $\beta$ with $b$, respectively. This means that all the individual photons are in improper mixed states. Thus, the putative ``Friends'' are nothing more than pointers P as discussed in the previous section, and as we have already seen, correlated pointers alone do not yield actual outcomes. Yet the authors seem to suggest that they are in proper mixed states. Specifically, the Appendix contains the following statement and characterization of the interaction: \bigskip

 ``{\it Depending on the state of the incoming photon}, the operation performed by Alice's friend transforms the overall state as follows:

$$ |h_a, \Psi ^- \rangle_{\alpha^\prime , \alpha}   = \frac {1}{\sqrt 2} [ |h\rangle _a   |h\rangle _{\alpha^\prime}
 |v\rangle _\alpha -  |h\rangle _a   |v\rangle _{\alpha^\prime}
 |h\rangle _\alpha]  \rightarrow \frac{1}{2} |h\rangle_a |v\rangle_\alpha,  $$
 
 $$ |v_a, \Psi ^- \rangle_{\alpha^\prime , \alpha}   = \frac {1}{\sqrt 2} [ |v\rangle _a   |h\rangle _{\alpha^\prime}
 |v\rangle _\alpha -  |v\rangle _a   |v\rangle _{\alpha^\prime}
 |h\rangle _\alpha]  \rightarrow \frac{1}{2} |v\rangle_a |h\rangle_\alpha  \eqno (S3) '' $$ 
 
 \bigskip

In the excerpt above, I have italicized the misleading phrase suggesting that the incoming photon {\it a} or {\it b} really is in a particular (but unknown) pure state, when that is not the case. In their equations (S3), which omit the entire entangled state of {\it a} and {\it b}, these transformations are presented as yielding well-defined outcomes for the source photons {\it a} and {\it b}, when in fact they do not. The actual transformation taking place upon the interaction with $\alpha$ is as follows:
 
 $$ {| \bar \Psi  \rangle}_{ab} |\Psi^-\rangle_{\alpha^\prime, \alpha}  \rightarrow 
 \frac{1}{2\sqrt2} [cos \frac{\pi}{8} (|h_a, v_\alpha , v_b\rangle + |v_a, h_\alpha, h_b\rangle)
 + sin \frac{\pi}{8} (|h_a, v_\alpha , h_b\rangle - | v_a, h_\alpha , v_b \rangle )]
     \eqno(8)  $$
    
   \bigskip
   
   This is clearly still an entangled state, so it is unambiguous that there has been no collapse or reduction as a result of the unitary interaction between $\alpha$ and the source state. Applying the transformation for Bob's ``Friend'' $\beta$ then yields the authors' equation (S7), an entangled state whose Bell-state properties can be confirmed via Alice and Bob's observable $A_1 = B_1$. ((S7) is just the final version of (8) including correlation of the $\beta$ states with $b$.)

     \smallskip
     
     Despite the fact that there has been no collapse or reduction as a result of the interaction of the PF*s with the source photons, the authors assert that an outcome occurred at that point. This sort of ambiguity about whether there is an outcome, how there could be an outcome, and even ``for whom'' there might be an outcome, is perhaps inevitable under the general assumption that quantum theory ``really'' is unitary-only, according to which even human beings must be quantum subsystems of much larger composite entangled states. The authors reveal this as their background assumption in discussing the original Wigner's Friend thought experiment, which involves a human being as the Friend. They say: 
   \smallskip
   
     ``Concurrently however, the friend does always record a definite outcome, which suggests that the original superposition was destroyed and Wigner should not observe any interference. The friend can even tell Wigner that she recorded a definite outcome (without revealing the result), yet Wigner and his friend's respective descriptions remain unchanged.'' 
     
     \smallskip
     
     The above says that ``the original superposition was destroyed'' but also that Wigner correctly describes the situation with a superposition: clearly a self-contradictory account. This kind of inconsistency is a feature of the ``shifty split'' of UOQM, in which collapse can be nothing more than a wholly mysterious, non-modeled process assumed to (somehow) occur based on observation by an ``external conscious observer.'' (Indeed the fact that ``external conscious observer'' cannot be defined constitutes the {\it reductio} nature of the WF experiment.) But there need be no inconsistency between the descriptions: if there really {\it were} non-unitary collapse as a result of the Friend's measurement, then Wigner would {\it not} be able to correctly model his system as remaining in a pure state. That is, had a definite outcome occurred along with collapse, the Friend would {\it not} be able to ``tell Wigner that she recorded a definite outcome without revealing the result'' and still have Wigner correctly describe his system by a pure state. Instead, Wigner's correct and experimentally verifiable description of the Friend and her system would be a proper mixed state, regardless of whether the Friend revealed anything (however uninformative) to Wigner or not. 
          
      Thus, the idea of the non-physical, mysterious collapse allegedly attending a ``measurement'' interaction leads the authors to the self-contradictory position of supposing that collapse ``must have occurred'' at the level of $\alpha$ and $\beta$ even though they themselves must deny collapse in order to obtain the correct final state (A7), and their own data show that no collapse occurred. The inability of UOQM to define what counts as a ``measurement'' or ``observation'' yielding an outcome -- because none can under UOQM -- is the ultimate source of the confusion here. It is perhaps worthwhile to note in this regard that Baumann and Wolf (2018) discuss the idea that an outcome only occurs ``relative to a subsystem'' (which they term``subjective collapse''). This is an attempt to preserve the idea that a subsystem defined as an ``observer'' may obtain an outcome that would theoretically lead to predictions inconsistent with those of the Wigner (super-observer) level, as noted above, yet such inconsistencies allegedly can never be empirically observed. Such a formulation appeals to the absence of a ``classical record'' in order to argue for the hidden nature of such inconsistencies. Yet the notion of a ``classical record'' itself depends on the unambiguous existence of measurement outcomes that, again, are never obtainable in the unitary-only formulation. 
      
      In any case, however, the claim that such inconsistencies must always remain hidden is untenable. A specific counterexample is provided in Kastner (2020b): an entangled subsystem degree of freedom defined as an ``observer,'' and to ``whom,'' based on that characterization, outcomes are attributed, could in principle yield an observable inconsistency with the outcomes of a ``super-observer.'' This counterexample demonstrates that attributing outcomes to entangled subsystems, even if designated as only ``relative'' to the subsystem and involving only ``subjective collapse,''  leads to empirical inconsistency signaling theoretical breakdown. In view of the importance of this issue, the counterexample will be repeated here. Let W and F be modeled as complex molecules with several excitable degrees of freedom subject to 2D state spaces.  Let  the system ``measured'' by F be labeled A, a spin-$\frac{1}{2}$ system, while F comprises 2 degrees of freedom labeled B and C.  Now, recall that according to UOQM, a ``measurement'' is nothing more than the establishment of a correlation between different degrees of freedom. In this account, the  F-level ``measurement'' of  A is  represented by a correlation between A and F's degree of freedom B. A is prepared in an equal superposition of outcomes `up' and `down'. C, F's communication degree of freedom, remains in its initial unexcited state $|0\rangle$ at this stage. According to the idea of a ``subjective collapse'' or ``relative outcome'' for F, after the ``measurement''  of A by B, A and F are either in the state 
      
      $$ |\Psi\uparrow\rangle_{FA} =   |\uparrow\rangle_A  \otimes |\uparrow\rangle_B \otimes |0\rangle_C\eqno(9a)$$
      or
       $$ |\Psi\downarrow\rangle_{FA} =  |\downarrow\rangle_A  \otimes |\downarrow\rangle_B \otimes |0\rangle_C\eqno(9b)$$
       \smallskip
       
\noindent with equal probability--i.e., they are in a mixed state. On the other hand, according to W,  A and F end up in a pure Bell state with F's communication degree of freedom C along for the ride:

        $$ |\Psi \rangle_{FA} = |\Phi^+ \rangle |0 \rangle_C = \frac{1}{\sqrt 2} \bigl ( |\uparrow\rangle_A  |\uparrow\rangle_B
        + |\downarrow\rangle_A  |\downarrow\rangle_B \bigl ) \otimes |0\rangle_C\eqno(10)$$
    
Now, let W subject the B+A system  to a measurement of the Bell observable for which the state $|\Phi^+\rangle$  is an eigenstate. W's experiment is accompanied by a signal to F as follows. An outcome finding the state
  $|\Phi^+\rangle_{FA}$  results in a photon being emitted to F to excite his communication degree of freedom C. According to W,  F should receive that photon for every run of the experiment.  This makes the inconsistency manifest, since according to F, his probability of receiving the photon is only 1/2  given his description of the situation by the mixed state (9a,b). Of course the practical logistics are nontrivial in carrying out this experiment, but nothing prevents it, in principle, under UOQM. Thus, the claim that one can attribute an outcome, however ``relative,'' to an entangled subsystem by calling it an ``observer'' leads to an in-principle observable inconsistency. It cannot be maintained that the resulting inconsistency is benignly ``hidden.'' Instead, theoretical breakdown results.

    The above considerations lead us to the Clauser-Horne-Shimony (CSH) inequality that is violated by the data obtained in this experiment. CSH (1969) originally formulated their inequality in terms of hidden variables, which are indeed forms of ``observer independent facts.''  But in the current context, which does not involve any hidden variables, the inequality arguably refers to the assumption that there were determinate measurement outcomes by the ``Friends,'' even though those putative outcomes did not collapse the superposition obtaining in the prepared Bell state. That is, the assumption of a ``relative outcome'' involving ``subjective collapse'' as defined by Baumann and Wolf (discussed above) plays the role of a hidden variable, since it is assumed to obtain even while the quantum state does not reflect that any outcome occurred (being still an entangled pure state). Since there are no other hidden variables assumed in this experiment, and the only nonstandard assumption is that the PF*s precipitated an outcome even without objective collapse or reduction of the quantum state, the violation of the inequality by the experimental statistics is properly taken as refuting that assumption. This is reinforced by the fact that the application of the quantum formalism required to match the experimentally observed results does not collapse the quantum state between the putative `measurement' by the PF*s and the detections by Alice and Bob. 
    
Of course, Proietti {\it et al} assume that the CHS inequality does not apply to their situation of putative ``relative'' outcomes. But the idea that their asserted outcomes only pertain to a local observer (and are therefore supposedly immune to the CHS inequality) is not supported by any mathematical or logical analysis. It's simply asserted as an {\it ad hoc} claim. It thus amounts to a subterfuge, a way of escaping the implications of the CHS inequality. Moreover, the claim that putative outcomes are restricted to a local ``observer'' (and therefore should be considered immune to the CHS inequality) is in fact falsified by the fact that such results could indeed be communicated among observers at the different levels, as shown herein and in Kastner (2020b). Therefore, the CHS inequality {\it does} apply to their assertion of outcomes both at the "Friend" and "Wigner" level, which is contradicted by quantum theory (which denies that any collapse occurred at the level of the PF*s) and falsified by the experimental findings.
       
     Thus, this experiment, rather than demonstrate an antirealist conclusion such as ``different observers irreconcilably disagree about what happened in an experiment,''  confirms what Feyerabend and Hughes pointed out decades ago: quantum theory (when taken as unitary-only) does not allow one to ascribe a fact of the matter, corresponding to an eigenvalue of a subsystem observable, to a subsystem of an entangled system.  A subsystem of an entangled system is therefore not an entity that can yield a well-defined ``fact'' corresponding to an outcome of an observable pertaining to that system. Something more is needed for an outcome (``fact'' ) to occur; specifically, physical collapse of the entangled (nonseparable) state to a single separable state. That has not occurred at the ``Friend'' level in this experiment. However, if the heralding photon were made into a pointer to the $h, v$  states, the final state (the authors' S7) would not be tenable even according to UOQM due to traditional decoherence arguments (i.e., putative entanglement with the degrees of freedom correlating the heralding photon to the relevant states). Of course, that account remains inadequate since there is still no justification for any definite outcome to arise under unitary-only dynamics.  For further discussion of this point, see Kastner (2014, 2020a, 2020b). 

The bottom line: Establishment of a correlating entanglement and verification of the correlating entanglement at the level of the ``Friend'' is not equivalent to ``outcome occurred'' or ``fact was obtained'' at the Friend level. The inequality (represented by Q above) expressing this erroneous assumption (P above) is thus violated by the experiment, which verifies the probabilistic predictions of quantum theory for the entangled state. Had reduction and a real outcome for the polarization observable occurred at the ``friend'' level (or even correlation of the heralding photon with the $h$ and $v$ states as mentioned above), there would be no ongoing entanglement at the level of Alice and Bob and that would be reflected in the statistics, which would confirm the loss of their prepared Bell state and its replacement by a mixed state (even if only an improper one under UOQM).

\section{So how do we get outcomes?} 

It is perfectly possible (beyond UOQM) to get a definite outcome, and of course we can all testify to the existence of definite outcomes in the lab. Rather than take the latter empirical fact as license to mis-attribute definite outcomes to systems in improper mixed states when that is logically contradictory (yet a practice that is endemic under the assumption of UOQM), a more consistent approach is to take our empirical experience as strong evidence that non-unitarity is real, and that this is why we get actual measurement results. The present author has been studying a formulation of quantum theory called the Relativistic Transactional Interpretation (RTI), which is based on the direct-action theory of fields. RTI contains quantitatively well-defined physical non-unitarity that does not involve any {\it ad hoc} change to quantum theory, and is empirically equivalent at the level of the Born probabilities to standard unitary-only quantum theory. While it is not the purpose of this paper to advocate any particular alternative formulation of quantum theory, the interested reader may consult Kastner (2018), (2020a) and (2021) for details. These references explain why RTI does not need to make any {\it ad hoc} change to the Schr\"odinger evolution as in the better known ``spontaneous collapse'' models such as that of Ghirardi, Rimini and Weber (1985). 

For present purposes, it may be noted that under RTI, decay rates $\Gamma(t)$ are actually the probability of collapse, involving photon emission, at any time $t$ for the usual situation in which an excited atom is surrounded by absorbers. Thus, any time there is a radiative process, collapse with respect to some observable has occurred. Under macroscopic conditions, the decay rates are very high and collapse is quite frequent, which constitutes the correspondence principle for the quantum/classical threshold under RTI. At the ``friend'' level of the Proietti {\it et al} experiment, the detection at the ``friend'' level involves only a photon whose state is uncorrelated to the observable of interest, so while there is collapse, it is not with respect to the observable of interest, so there is no outcome established for that observable. However, collapse occurs with respect to the relevant observable at the ``Alice and Bob'' level, which is why Alice and Bob find definite outcomes confirming their Bell state. In Wigner's original thought experiment, without restriction to UOQM and under a direct measurement by the macroscopic Friend of the system presented to him, collapse could easily occur at that stage. (What counts as ``macroscopic'' and why that level generally precipitates collapse is discussed in Kastner (2018)).  Then, even if the Friend ``merely'' signaled to Wigner that he saw a result without revealing the result, the system would still have been collapsed, and Wigner would find his system in a proper mixed state, not a pure state.  Thus, we have to be careful about assuming that such a signal (``I see a result but I'm not revealing which'') preserves entanglement. In fact it does not if reduction actually occurred at that point. On the other hand, if entanglement is preserved beyond that point, then nobody really ``saw a result,'' since there could be no well-defined result (as Feyerabend and Hughes point out). Thus any portrayal of a ``Friend'' as really seeing a result but not disclosing it and yet continuing on as a component of an entangled state is erroneous, since no definite outcome can occur under those circumstances.\footnote{Zukowski and Markiewicz (2021) argue that there is no measurement outcome at the ``Friend'' level in this experiment using decoherence arguments. However, decoherence alone is not sufficient to establish a measurement outcome; this is implied by Hughes' argument and is discussed in detail in Kastner (2014) and references therein. See also Kastner (2020a).} 

\section{Conclusion}

It has been argued that the observed violation of the relevant CHS inequality, rather than supporting claims such as ``observers see irreconcilable facts'' demonstrates that it is a mistake to attribute an outcome or `fact' of the matter concerning any specific observable to a system that is in an improper mixed state, such as the PF*s in the Proietti {\it et al} experiment.

The authors do not consider this interpretation of the violation of the CHS inequality, because they do not question their claim that the PF*s did find an outcome. Yet, as we reviewed above, it is unambiguously incorrect to assert that the PF*s found an outcome, since that assumption violates quantum theory itself: a degree of freedom in an entangled pure state, like the PF*, is in an improper mixture that cannot be described by a definite but unknown observable eigenstate. The observed violation of the CHS inequality expressing the incorrect assumption that an outcome occurred at the PF* level confirms this elementary fact. 

Another way to see that this is an error is to consider one of the electrons in a two-electron Bell state. Yes, the electron's state is correlated to its partner, but that is not justification for attributing any measurement outcome, i.e., value of an observable, to the entangled electron. This mistake is made by the authors when they say: \bigskip

``Recalling from Eq. (S4) how the friends' measurement results are encoded in their polarisation states\ldots ''
\bigskip This is no more correct than asserting that a Bell state like

$$\Psi\rangle = \frac{1}{\sqrt 2} [|z\uparrow\rangle|z\downarrow\rangle - |z\downarrow\rangle|z\uparrow\rangle] \eqno (11) $$

\bigskip
 \noindent expresses an `encoding of measurement results.'  Yes, the electron's state is correlated to its partner, but that does not mean that any `information has been extracted' nor that there can be any value of Z-spin -- i.e., no outcome -- attributed to the electron.
 
     Finally, the point of this paper is not to quibble about whether the internal photons $\alpha$ and $\beta$ count as ``observers'' or not, since UOQM is utterly helpless to say anything about what an ``observer'' could possibly be.  If an ``observer'' is a system that can yield an outcome, then no system, not even a human being, can give rise to a well-defined outcome under UOQM. As we reviewed in Part 1, the mere copying of a quantum number pertaining to one degree of freedom by another degree of freedom (the definition of ``observer'' used by the authors) does not qualify as yielding an outcome, despite the language about ``extracting information.'' (This is just the measurement problem facing UOQM again.) Thus, our routine experience of definite outcomes remains at odds with UOQM. Nevertheless, it is quite clear from the data correctly predicted by the final uncollapsed state (the authors' (S7)) that there is no collapse at the level of the PF*s. The fact that they copied quantum numbers in a superposition, i.e., acted as pointers, does not establish any fact of the matter; there is no warrant to claim that they ``observed an outcome''  since no outcome is possible according to quantum theory itself. That is, quantum theory itself does not permit the interpretation of the improper mixed state of a subsystem of a composite pure state as a situation in which an outcome occurred,``relative'' or not. Therefore, it is not tenable to claim that there was a measurement outcome at the level of $\alpha$ and $\beta$. The ``heralding'' photons $\alpha^{\prime}$ and $\beta^{\prime}$ merely signal that a unitary interaction has occurred at that point, not that there was any outcome. The observed violation of the relevant CHS inequality, rather than indicating that there are irreconcilable facts among observers, demonstrates the falsity of the assumption that there was a ``fact'' (outcome) at the internal Friend level. Thus, rather than a test of local observer independence (as stated in the title), this experiment is a test of the assumption that there is a measurement outcome based only on a non-decohering unitary correlation among photons. That assumption fails the test.
     
\newpage

\section{References}

Baumann, V. and Wolf, S. (2018). ``On Formalisms and Interpretations,'' {\it Quantum 2}, 99.

Bong, K.-W. et al (2020). ``A strong no-go theorem on the Wigner?s friend paradox,'' {\it Nature Physics, Vol.16}, 1199-1205.

J. Clauser, M. Horne, A. Shimony, and R. Holt (1969). {\it Phys.Rev. Lett.23}, 880.

Feyerabend, P. K.  (1962). ``On the Quantum Theory of Measurement.'' In Korner S. (ed.) (1962). {\it Observation and Interpretation in the Philosophy of Physics}. New York: Dover, pp. 121-130.

Frauchiger, D. and Renner, R. (2018). ``Quantum theory cannot consistently describe the use of itself,'' {\it Nature Communications 9}, Article number: 3711.

Ghirardi, G.C., Rimini, A., and Weber, T. (1986). {\it Phys. Rev. D 34}, 470.

Hughes, R. I. G. (1989). {\it The Structure and Interpretation of Quantum Mechanics}. Cambridge: Harvard University Press. 

Kastner, R. E. (2012). {\it The Transactional Interpretation of Quantum Mechanics: The Reality of Possibility.} Cambridge: Cambridge University Press.

Kastner, R. E. (2014). `` 'Einselection' of pointer observables: the new H-theorem?'' {\it Stud. Hist. Philos. Mod. Phys. 48}, 56-8.

Kastner, R. E.  (2018). ``On the Status of the Measurement Problem: Recalling the Relativistic Transactional Interpretation,'' {\it Int'l Jour. Quan. Foundations 4}, 1:128-141.

Kastner, R. E. (2020a). ``Decoherence in the Transactional Interpretation,'' {\it Int'l Jour. Quan. Foundations 6}, 2:24-39. 

Kastner, R. E. (2020b). ``Unitary-Only Quantum Theory Cannot Consistently Describe the Use of Itself: On the Frauchiger-Renner Paradox,'' {\it Foundations of Physics 50}, 441-456.

Kastner, R. E. (2021). ``The Relativistic Transactional Interpretation and The Quantum Direct-Action Theory,'' preprint. https://arxiv.org/abs/2101.00712. (This material is excerpted from the forthcoming second edition of Kastner (2012).

Proietti, M. et al (2019). ``Experimental test of local observer independence,'' {\it Science Advances}, 
Vol. 5, no. 9. DOI: 10.1126/sciadv.aaw9832 

Zukowski, M. and Markiewicz, M. (2021). ``Physics and Metaphysics of Wigner's Friends: Even Performed Premeasurements Have No Results,'' {\it Phys. Rev. Lett. 126} (13).

\end{document}